\documentclass[12pt]{amsart}
\usepackage[english]{babel}
\usepackage{latexsym,amsmath,amsfonts,amscd,amssymb,latexsym}
\usepackage{graphicx}
\usepackage{xcolor}
\newtheorem{theorem*}{Theorem}

\newtheorem{proposition*}[theorem*]{Proposition}

\newtheorem{corollary*}[theorem*]{Corollary}

\theoremstyle{remark}

\newtheorem{remark*}[theorem*]{Remark}

\newtheorem{note*}[theorem*]{Note}

\def\bitcoinA{%
  \leavevmode
  \vtop{\offinterlineskip %\bfseries
    \setbox0=\hbox{B}%
    \setbox2=\hbox to\wd0{\hfil\hskip-.03em
    \vrule height .3ex width .15ex\hskip .08em
    \vrule height .3ex width .15ex\hfil}
    \vbox{\copy2\box0}\box2}}

\begin{document}

\title{\centerline{Ping-Pong Swaps} 
\smallskip 
\tiny{Peer-to-peer crosschain swaps 
without\protect \\ escrow nor trusted third party}}

% \date{23 November 2022}

\author{Cyril Grunspan}
\address{L{\'e}onard de Vinci, P{\^o}le Univ., Research Center, Paris-La D{\'e}fense, France}
\email{cyril.grunspan@devinci.fr}
\author{Ricardo P\'erez-Marco}
\address{CNRS, IMJ-PRG,  Paris, France}
\email{ricardo.perez.marco@gmail.com}
\address{\tiny Author's Bitcoin Beer Address (ABBA)\footnote{\tiny If you find this article useful, 
you can send some anonymous satoshis to support our research at the pub.}:
% \newline
% {}\indent 
1KrqVxqQFyUY9WuWcR5EHGVvhCS841LPLn} 
\address{\includegraphics[scale=0.3]{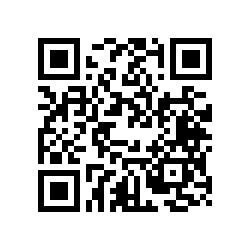}}

\begin{abstract} 
We propose Ping-Pong Swaps: A secure pure peer-to-peer crosschain swap mechanism of tokens or cryptocurrencies that does not require escrow 
nor an intermediate trusted third party. The only technical requirement is to be able to open unidirectional payment channels in 
both blockchain protocols. This allows anonymous cryptocurrency trading without the need of a centralized exchange, nor DEX's in 
DeFi platforms, nor multisignature escrow systems with penalties. Direct peer-to-peer crosschain swaps can be performed without 
a bridge platform. This enables the creation of non-custodial exchanges 
and also a global peer-to-peer market of pairs of tokens or cryptocurrencies. 
Ping-pong swaps with fiat currency is possible if banks incorporate simple payment channel functionalities. Some immediate 
applications are simple and fast rebalancing of Lightning Network channels, and wrapping tokens in smartchains.
\end{abstract}

\maketitle

\section{Introduction}
Satoshi Nakamoto's foundational article \cite{N08} describes Bitcoin protocol as a peer-to-peer payment protocol.
After Bitcoin's invention, many other cryptocurrencies where developed following the Bitcoin code or ideas. Bitcoin 
is also the first programmable form of money and smartcontracts of different sorts can be implemented using its scripting language.
Some years later,  smartchains were build, specifically dedicated to smartcontracts, and it was easy to 
create tokens for various purposes using smartcontracts (see also \cite{Frei} for an earlier proposal of 
extension of the Bitcoin protocol), like 
for example ERC-20 tokens in the Ethereum 
network. These tokens are useful for ICO's but also some of them represent stable coins, like for example Tether 
whose value is tied to the USD. Most of the trading 
occuring in cryptocurrency exchanges involves trading pairs A/B of two cryptocurrencies or tokens. These trades in exchanges 
are centralized are processed in the internal balances of the exchanges and require the trust on the exchange, and thus 
have a counterparty risk (for example the risk of fractional reserve and insolvency of the exchange). Our goal 
is to extend Nakamoto's peer-to-peer payment philosophy to the global cryptocurrency market.

When both tokens A and B are in the same smartchain, one can devise smartcontracts that allow a decentralized trade of the pair A/B 
through the interaction with the smartcontract. This is the basic principle of Decentraliced Exchanges (DEX's), like Uniswap 
in the ETH network, or Cake in the BSC network. These are fundamental pieces of the DeFi technology that is being build.

When both tokens or cryptocurrencies A and B do not belong to the same blockchain, the situation is more complex. Some 
cryptograhic solutions of ``Atomic Swaps'' (\cite{Atomic}) have been proposed in certain blockchains 
with specific properties (see for example
the Decred project \cite{decred}). Atomic swaps required more cryptographic functionalities than the solution of ping-pong swaps 
proposed here. More precisely, they need also a hashlock mechanism which is not needed for the solution we present here.

Other peer-to-peer protocol exchanges work in the base of an escrow system 
and penalties in case of dishonest behavior, as for example Bisq (\cite{Bisq}) or Hodlhodl (\cite{hodlhodl}).
Also a ``bridge industry'' is being developed for crosschain swaps (Ren, Anyswap, Axelar, Thorchain, etc). 
Many of these projects aim to be decentralized but so far they do not achieve the decentralization quality that one 
expects from Bitcoin standards.

The goal of this article is to propose a very simple procedure that allows peer-to-peer decentralized swaps without escrow 
nor trusted third party involved (and, of course, no mutual trust required between both parties). 
The only important (but mild) technical requirement is that in both blockchains we can open simple
unidirectional payment channels. An unidirectional payment channel can be open if we can create 2-2 multisignature transactions 
and we dispose of timelock scripts.

\section{Payment channels}

\subsection{General payment channels.}
Payment channels allow off-chain transactions that are instant and secure. They also allow micropayments and have no commission
cost, except for the opening or closing of the channel on-chain. They require smartcontracts with a 2-2 multisignature and a timelock 
script. They can be unidirectional, i.e. allowing the exchange of the cryptocurrency in only one direction, or bidirectional. Unidirectional 
payment channels are much simpler than bidirectional ones that require  
a constant monitorization of the blockchain. This is not necessary in the case of unidirectional payment channels. Payment channels 
were first proposed by Satoshi Nakamoto himself in some private emails, and  J. Spillman proposed a simple unidirectional payment channel 
that could be implemented after Segwit. Later, other sorts of more complex payment channels 
were proposed  (see \cite{channel}). The Lightning Network that is currently being implemented, is build using the 
composition property of bidirectional Poon-Dryja payment channels, as described in the Lightning Network article \cite{Poon-Dryja}. 
For our purposes we only need the simpler Spillman's unidirectional channels, that do not need a hashlock scrypt. But of course, the same
protocol will work with bidrectional payment channels or their composition, and in particular we can use in the Bitcoin blockchain the 
Lightning Network currently implemented. We describe in the next section
how to build the simpler unidirectional payment channels in the Bitcoin network. In other smartchain with a 
rich scripting programming languages, one can build smartcontracts 
that simulate multisignatures, timelocks and unidirectional payment channels. We will not discuss the specifics for each smartchain or 
token type. With a rich scripting language the task is in general much easier than in the Bitcoin network. For example, in the 
Ethereum network, the goal of the Raiden project \cite{Raiden} is to create a Lightning Network for tokens in the ETH blockchain 
and they have developped payment channels.

\subsection{Unidirectional payment channel in Bitcoin.}
We describe how to build an unidirectional payment channel from Alice to Bob in the Bitcoin protocol (or similar). 
Alice wants to pay Bob. 
Alice and Bob agree on the 
total amount of bitcoins that will be locked in a 2-2 multisignature address with two outputs to addresses controlled by each one of them. 
Bob provides a signed 
transaction with the total amount and a timelock, so that Alice can recover the totality of her funds if the channel is 
not closed before some predetermined time $t_0$ (in blockchain time). Then Alice sends the total 
amount of bitcoins to this 2-2 multisignature address.
This on-chain transaction opens the payment channel. Now, each time Alice wants to pay Bob, she only needs to sign a 
1-2 signature new transaction which transferts the new balance to the output addresses.
Bob can cash-out anytime by providing the second signature and broadcasting the transaction. This action will close the payment channel.
But Bob waits for more payments from Alice. In any case, Bob must close the channel before $t_0$ or Alice will be 
able to recover all the funds. 

With this procedure, Alice can make instant payments or micro-payments with no fee involved that are done off-chain by simple
data communication from Alice to Bob. The computational cost is only 
formatting the transaction with 1 of 2 signatures. The ECDSA (or Snorr) signature requires only a small fraction of a second.

\section{Ping-pong swaps.}

We explain in this section ping-pong swaps with a very specific example using the Bitcoin and Litecoin networks, 
from which the general procedure is clear. The code of Litecoin 
is a fork of the code of Bitcoin and has incorporated most of the script functionalities from Bitcoin, thus we can also 
open in the Litecoin network Spillman unidirectional payment channels.

\subsection{Preliminary setup and opening channels.}
Imagine now the situation where Alice wants to buy from Bob a cryptocurrency. For example, 
Alice has BTC and wants to buy LTC from Bob. In both of these 
two blockchains we can create unidirectional payment channels as described in the previous section. 
First they agree on an amount and a price for the 
trade of the pair BTC/LTC. Suppose that Alice wants to spend 1 BTC and they agree on the price 
$1 \ \text{LTC}= 0.003521 \ \text{BTC}$ (current exchange 
rate at the moment). This means that Bob has to provide $284 \ \text{LTC}$ to Alice. 

In the BTC network, Alice opens an unidirectional channel with Bob for the total amount of $1$ BTC. In the LTC network, Bob
opens an unidirectional channel with Alice in the amount of $284 \ \text{LTC}$. They both agree on timelocks (1 hour 
for example, enough time to avoid double spends with the timelocked recovery transactions).

They also agree on the granularity of the swap, which can be, for example, of 0.1\% of the total amount. This is some small fraction 
of the total amount that is not worth cheating from any of the parties involved in the trade.

\subsection{Ping-pong micropayments.} After this setup, Alice and Bob engage in a ``ping-pong'' sequence of micropayments.
First Alice pays $0.1\%$ of the total amount (that is, $100.000$ satoshis) to Bob through the Bitcoin payment 
channel. Once Bob has received the 1-2 signed transaction from Alice, he pays $0.2\%$ of the total amount of LTC ($0.568 \ \text{LTC}$ to 
be precise) to Alice through the LTC payment channel. Then, once Alice has received Bob's payment, she sends $0.2\%$ of 
the total amount of BTC, then Bob sends again $0.2\%$ of LTC, 
and so on. In the last transaction Alice will send only $0.1\%$. In this example, Alice will perform a 
total of $501$ payments and Bob $500$ payments. Once
they both transfer the total amount, they can close the channels and recover the funds in an address of their own.

Obviously, all these micropayments are done by a software wallet designed for this swap purpose. In the example, the 
total transaction is completed after $1001$ signatures, thus in less than a second if the channel payments (signature) 
take under a millisecond. In general, the processing time is inversely proportional to the granularity of the swap.

\subsection{Cheating.}
The important observation, is that the only way that Alice or Bob can cheat is by interrupting the ``ping-pong'' 
transactions and closing their receiving channel, pocketing the granularity of the payment ($0.1\%$ in our example). 
The other party will then close the channel to secure the payment already received. The granularity can be adjusted to have a 
small enough value to desincentivize any cheating. If the granularity is smaller than the fees for opening and closing the channels, 
there is no profit for the cheater (although he saves one granularity).

\section{Implementation.}

In principle, this can be easily implemented as a mobile app having some functionalities of a software wallet in a computer or 
mobile phone. Once
Alice and Bob agree on their trade, the app provides the necessary information to the other party to open the channels, and each one 
opens its unidirectional channel. Both apps communicate directly for the ``ping-pong'' of transactions, and finalizing the swap
by closing the channels. This should be almost instantaneous, except for the waiting time of on-chain transaction to open and close 
the channels. If both Alice and Bob have open payment channels in the Lightning Network in BTC, they can use them as one of the channels.

\section{Non-custodial exchanges (NCEX).}

In order to find counterparts, the app can connect to a server where people can post their trade offers. 
The matching can be done automatically or manually. The advertisement of the offers involves then a 
centralized server, but only for posting and reputation purposes. 
In this way we can build a centralized urely cryptocurrency Non-Custodial EXchange (NCEX), 
where the users have control of their funds all the time. This type of exchange limits 
the liability of the exchange operator since the exchange cannot be hacked in order to access its wallets. Also it is 
interesting to note that if the exchange charges a minimal fee higher than the granularity and the fees for opening the 
payment channels, there is no point in cheating on the granularity of the ping-pong swap.

In the implementation of non-custodial exchanges, there is the possibility of doing the ping-pong swaps through the 
exchange. The added value of the exchange could be, for example, in having a reputation system and database of trusted dealers.
Alice and Bob can be users of the exchange and open an important payment channel with the exchange for the total amount 
of BTC that Alice wants to trade and the total amount of LTC that Bob wants to trade. Then, once the orders are matched, 
two ping-pong swaps can be performed through the exchange by the composition of the BTC payment channel from Alice to the exchange and 
from the exchange to Bob, and the LTC channels from Bob to the exchange and the exchange to Alice. The dynamics of the composed 
ping-pong swaps is easy to figure out. First Alice send his initial $0.1$\% payment to the exchange, then the exchange forwards it 
to Bob. Once Bob receives the first micropayment it sends his first $0.2$\% LTC micropayment to the exchange that forwards it 
to Alice, etc. If Alice has a large sell order of BTC that can only be absorbed by several buyers, in this way she only needs 
to open two channels with the exchange (thus only two on-chain fees). Thus this type of procedure saves important on-chain fees although
is not P2P having the exchange as middle-man, but it is still a trustless procedure.

\section{A decentralized global market.}

One can imagine a better decentralized exchange system using a  
decentralized communication protocol similar to the one used in the Bitcoin network. In this setup, 
offers can be propagated flooding the network, as is done in the broadcasting of transactions in the Bitcoin network.
Each app is a communication node, connects to 
8 or more neighbours, and relays the ``mempool'' of offers for each trading pair A/B. This creates an international decentralized 
bid-ask order book for each trading pair. Another possibility is to build a federation of Nostr relays where the offers 
are posted and matched by the clients (\cite{Nostr}).

In this way we can build a global purely decentralized exchange and worldwide global market, 
which is uncensorable and with no centralized structure. 
Properly implemented, this glogal market offers anonymous trading, with almost
no fees. The classical exchange and broker industry will still be useful for crypto/fiat pairs, or for those pairs of  
cryptocurrencies that involve one whose 
blockchains does not allow payment channels.

\section{Application: Wrapping cryptocurrencies into other chains.}

Non-native cryptocurrencies can be used in other smartchains if they are wrapped 
into native tokens. For example, bitcoins are wrapped into the Ethereum blockchain as tokens WBTC, in the Binance 
Smart Chain as BTCB, in RSK network (\cite{RSK}) as R-BTC, or in Liquid network (\cite{Liquid}, \cite{sidechains}) as L-BTC. 
The procedure of wrapping can be more or less decentralized or
elaborate depending on the chain. In Ethereum and the Binance smartchains, the tokenization is done by a trusted third party,
a company or an exchange.
In most of the situations, the tokenization can be done through a centralized exchange.
In Liquid and RSK the procedure can be more decentralized, but requires an important number of confirmations of the transaction 
in the Bitcoin network (over 100 for RSK and Liquid, thus more than 16 hours). 
The wrapping procedure may also involve extra fees other than the transaction fees. 

With ping-pong swaps we can simplify the 
procedure, taking advantage of anyone willing to unwrap tokens, with a ping-pong swap between the native and wrapped tokens. 
For example, Alice having native BTC, can initiate a ping-pong swap with Bob having wrapped BTC. Alice
opens a payment channel with Bob in the BTC network, and Bob one to Alice in the smartchain. If the smartchain has
fast confirmation times, the procedure takes the time of the confirmations for opening and closing the BTC channel.
For example, we could wrap BTC and obtain R-BTC in RSK with a ping-pong swap in 4 confirmations (2 confirmations 
for opening and closing the channel 
is enough for security purposes, see \cite{GPM}) instead of 100 confirmations required by
RSK's powpeg procedure.

\section{Application: Rebalancing Lightning Network channels.}

One of the most common problem of the LN as is being implemented is the rebalancing of payment channels. The channels used 
are all bidirectional. Unbalanced channels become unidirectional and their routing capability is impaired. Several solutions 
have been proposed to balance LN payment channels. By circulary, if we can find a convenient circular composition of channels 
that allows the rebalancing. Since the general flow of payments is from customers to vendors, this is not a realistic 
general solution. Other solutions like splines involve on-chain transaction, thus on-chain fees and waiting time.
Here we propose a solution using ping-pong swaps using a secondary chain that has fast confirmations and cheap fees.

In case Alice and Bob have an open LN payment channel that they want to rebalance by sending funds from Alice 
to Bob, they can run a simple ping-pong swap. They will use the existing bidirectional LN channel as Alice payment channel, 
and Bob will open an unidirectional channel to Alice in the Bitcoin network or another blockchain (in this last case, they need 
to agree first on the conversion rate). They will proceed exactly as before. 
This procedure only needs two on-chain transactions for opening and closing the channel from Bob to Alice.

Notice that this allows to rebalance your Bitcoin LN channel using another cryptocurrency or token. If we use one 
that has fast confirmation time in its blockchain (as the Ethereum network), the whole procedure of rebalancing can be 
performed in seconds. If we use one that has cheap transaction fees, the cost will be small.

\section{Fiat swaps}
Ping-pong swaps can be performed with cryptocurrencies/fiat pairs if commercial banks incorporate the payment channel technology.
In their centralized setting, this is very simple. Banks need to allow high frequency micropayments between client
accounts. They need to have API's that allow the interaction of the users cryptowallet with their bank account in order to 
check the dynamics of the ping-pong micropayments. Then you can setup for example a BTC/USD ping-pong swap. Both Alice and Bob 
need to be customers of the bank, and Alice will open a payment channel to Bob's account internal to the bank. The app will 
perform the ping-pong  swap checking with the APIs the bank account balance and ordering the micropayments.

Incorporating this technology, banks can become cryptoexchanges. They will only be custodial of the fiat funds. The users
keep at all times control of their cryptocurencies with their private keys.

\bigskip
% \bigskip

\textbf{Remark.} Without doubt, the core idea of making first small trades in order to build confidence with an 
unknown dealer goes back to the origins of trade. Some traders use this procedure when buying from early miners 
new cryptocurrencies not listed in exchanges. The contribution of ping-pong swaps is the combination of this old folklore 
procedure with micropayments through payment channels. Not surprinsingly, this simple idea appeared before somewhere else. 
After writing the article, it was pointed out to the authors the blog post by Micah Zoltu \cite{Zoltu} where a ping-pong 
swap of the pair ETH/ZEC is described. 

% \bigskip

\end{document}